# On the relevance of gravitational self-force corrections on parameter estimation errors for extreme-mass-ratio inspirals

E A Huerta[1] and Jonathan R Gair[1]

[1] Institute of Astronomy, Madingley Road, CB3 0HA, Cambridge, UK

Email: eah41@ast. cam.ac.uk, jgair@ast.cam.ac.uk

**Abstract**. It is not currently clear how important it will be to include conservative self-force (SF) corrections in the models for extreme-mass-ratio inspiral (EMRI) waveforms that will be used to detect such signals in LISA (Laser Interferometer Space Antenna) data. These proceedings will address this issue for circular-equatorial inspirals using an approximate EMRI model that includes conservative corrections at leading post-Newtonian order. We will present estimates of the magnitude of the parameter estimation errors that would result from omitting conservative corrections, and compare these to the errors that will arise from noise fluctuations in the detector. We will also use this model to explore the relative importance of the second-order radiative piece of the SF, which is not presently known.

## 1. Introduction

The cores of most galaxies are expected to host massive black holes (MBHs) which are surrounded by clusters of stars that contain large numbers of compact stellar remnants. As a result of mass segregation, these compact objects (COs) may be captured by the central MBH and eventually inspiral under the influence of radiation reaction from the emission of gravitational waves (GWs) [1]. The phase of the GWs from these extreme-mass-ratio inspirals (EMRIs) will contain information that will be used to map the spacetime exterior to the large body and explore the response of the horizon to tidal forces. Furthermore, since LISA will be able to use this information to measure the BH's mass, spin, and quadrupole moment to fractional accuracies of $10^{-3}$[10], it will be possible to test the predictions that general relativity makes about black hole solutions.

Nonetheless, the detection of EMRIs will be a challenging endeavor. The expected GW signal will be buried by instrumental noise and a foreground of galactic white-dwarf binaries. In order to dig out the signal from the noise we will need to use matched filtering. However, this extraction mechanism will only be successful as long as the theoretical waveform template is accurate, i.e., its phase should remain accurate to one cycle over the $M/m$ cycles of the waveform generated while the CO emits GWs in the strong curvature region. Black Hole Perturbation Theory (BHPT) is the appropriate framework to develop accurate templates using the mass ratio $\eta$ as a small expansion parameter. However, first order radiative BHPT in a Kerr background is described by the Teukolsky formalism and such waveforms are very computationally expensive.

An alternative approach to EMRI modeling is to develop approximate models that capture the main features of true inspirals but which are computationally inexpensive. We have used a numerical kludge (NK) waveform model that combines an exact particle trajectory with a flat space-time wave-emission formula. This prescription captures the main features of the waveform accurately as the overlap between Teukolsky-based and NK waveforms is greater than 0.95 over a considerable portion of the parameter space [2]. Additionally, we have augmented our NK model by radiative SF corrections derived from perturbative calculations, and conservative SF corrections derived by comparison to post-Newtonian (PN) results in a physically consistent way. In these proceedings we aim to shed some light on the importance of the conservative SF piece for waveform modeling, both in terms of source detection and parameter estimation. We introduce our improved NK model in section 2. Section 3 presents a brief summary of signal analysis; the results are summarized in section 4; conclusions and future improvements are presented in section 5.

## 2. Source modelling

Since the system's extreme-mass-ratio $\eta \sim 10^{-6}$ guarantees that gravitational back-reaction effects occur on timescales much longer than any orbital timescale, we will assume that the inspiralling CO instantaneously follows a Kerr space-time geodesic. We shall restrict our attention to circular-equatorial orbits for which the inclination angle and the eccentricity remain constant [3]. We include the first order radiative piece of the SF to slowly evolve the parameters of the geodesic using a consistent prescription for the energy and angular momentum fluxes, i.e., circular orbits remain circular under radiation reaction [4]. Because the "circular goes to circular rule" guarantees that we only need specify either the energy or the angular momentum to evolve a circular orbit, we choose to evolve the geodesic parameters using an angular momentum flux matched to Teukolsky-based evolutions. Under this prescription, the radial coordinate $p(t)$ is evolved as follows,

$$\frac{dp(t)}{dt} = \frac{dp(t)}{dL_z}\frac{dL_z}{dt}. \tag{1}$$

The conservative piece of the SF has two parts. One is oscillatory and averages to zero whereas the second component affects the phasing of the waveform over time. We take account of this effect by rewriting the orbital frequency as follows,

$$\frac{d\phi}{dt} = \left(\frac{d\phi}{dt}\right)_{geo}[1+\delta\Omega]. \tag{2}$$

Equation (2) includes the phase derivative for a geodesic, "*geo*", and the frequency shift $\delta\Omega$ that should be computed using the SF formalism. These shifts are known only for a point-like particle moving on a circular orbit around a Schwarzschild black hole in a particular gauge [5]. In order to make progress, we have used conservative corrections in the PN framework at 2PN order which include spin corrections, finite mass contributions and radiative SF expressions [6], [7] to augment the kludge. By forcing two asymptotic observables, namely, the orbital frequency and its first time derivative, to be consistent with the PN results in the weak field, we can identify coordinates between the two formalisms and find the conservative pieces missing in the kludge [8].

Having derived these various corrections, we can assess their influence on the parameter estimation accuracy which LISA observations are likely to achieve. We build our waveform by using a flat-space quadrupole wave-generation formula applied to the trajectory of the inspiralling CO in Boyer-Lindquist coordinates, which are identified with spherical-polar coordinates in flat space-time (see [8] and references therein for details). Following [9] we implement the LISA response function to write the NK waveform as

$$h_\alpha(t) = \frac{\sqrt{3}}{2D}\left[F_\alpha^+(t)A^+(t) + F_\alpha^\times(t)A^\times(t)\right], \tag{3}$$

where $\alpha$ = I,II refers to the two independent Michelson-like detectors that constitute the LISA response at low frequencies, $F_\alpha^\pm(t)$, $A_\alpha^\pm(t)$ are the antenna pattern functions and the polarization coefficients, respectively. In our studies we will consider an orbital period of one year. We also include Doppler phase modulations in the detector response (see [8] for further details).

*Assessing the influence of the second order radiative piece of the self-force in the non-spinning limit.* Using SF results we have also built asymptotic observables within the SF formalism. However, by construction in the SF gauge, we are missing a contribution from the second order part of the angular momentum flux. We can compute this correction using the same PN results used to correct the kludge and then assess the importance of the second order radiative piece of the SF by computing the difference between an evolution using only an angular momentum flux matched to Teukolsky based-evolutions and one that also includes the second order correction [8].

**3. Data analysis**

We will explore a 10D space comprising four intrinsic parameters, namely, the SMBH's mass and spin, the mass of the CO, and the initial radius of inspiral, and six extrinsic parameters, namely, the initial phase of inspiral, two angles to describe the position of the source in the sky, two angles to describe the orientation of the SMBH's spin, and the distance to the source. We aim to reconstruct the most probable value of the parameter $\theta = \{\theta_1,...,\theta_{10}\}$ of the source and estimate their respective errors.

For large SNR, the expectation value of the errors $\Delta\theta^i$ is given by

$$\left\langle \Delta\theta^i \Delta\theta^j \right\rangle = (\Gamma-1)^{ij} + O(\text{SNR})^{-1}, \tag{4}$$

where $\Gamma^{ij}$ is the Fisher information matrix which can be written approximately as

$$\Gamma_{ab} = 2\sum_\alpha \int_0^T \partial_a \hat{h}_\alpha(t) \partial_b \hat{h}_\alpha(t) dt, \tag{5}$$

where $\hat{h}_\alpha(t)$ is a noise-weighted waveform that is computed using the total LISA noise, which comprises instrumental noise, confusion noise from short-period galactic binaries, and confusion noise from extragalactic binaries [10].

3.1. *Parameter estimation errors.* To estimate errors using the inverse Fisher Matrix we used fixed values for the intrinsic parameters and the distance to the source, $D = 1\text{Gpc}$, and carried out a Monte Carlo simulation over values of the remaining five extrinsic parameters. Note that Fisher Matrix errors are SNR dependent, but we quote results at fixed SNR=30. We renormalize the results after computing the SNR at $D = 1\text{Gpc}$ via the expression

$$\text{SNR}^2 = 2\sum_{\alpha=\text{I,II}} \int_{t_{init}}^{t_{LSO}} \hat{h}^2{}_\alpha(t) dt. \tag{6}$$

3.2. *Model errors.* Apart from errors that arise due to detector noise, we need to acknowledge that our NK is only approximate and hence the kludge waveform that is the best-fit to the data may have different parameters to the true waveform, introducing another parameter error. We shall use the formalism introduced in [11] to estimate theoretically what error would result from omitting conservative corrections from the waveform model, or by using a model with incomplete conservative corrections. We shall take the "true" waveform, $h_{\text{GR}}$, to be the NK waveform including all conservative pieces at 2PN order for the case of Kerr EMRIs. For Schwarzschild EMRIs, the true waveform will include PN second order radiative corrections in the angular momentum flux in addition to the first order flux matched to Teukolsky based-evolutions. We then estimate the model errors by searching for $h_{GR}$ using approximate templates, $h_{AP}$, that include none or part of the conservative/radiative corrections using the approximation [11]

$$\Delta_{th}\theta^i \approx \left(\Gamma^{-1}(\theta)\right)^{ij}\left(\underbrace{[\Delta\mathbf{A}+i\mathbf{A}\Delta\mathbf{\Psi}]e^{i\mathbf{\Psi}}}_{\text{at }\theta}\Big|\partial_j\mathbf{h}_{AP}(\theta)\right). \quad (7)$$

In equation (7) $\theta$ is the best fit, and the left hand side entry of the inner product corresponds to the difference between the "true" and approximate waveforms in an amplitude, $\mathbf{A}$, phase, $\mathbf{\Psi}$, decomposition. Notice that model errors are both noise and SNR independent.

## 4. Results

We have assessed the importance of including conservative corrections by estimating how long all the conservative pieces take to contribute one cycle to the GW phasing. Our results suggest that it may be necessary to include conservative corrections in EMRI models, but it may not be necessary to go beyond 2PN order. Similarly, we have also found that the missing second order radiative piece from the SF formalism appears to be relatively unimportant for the gravitational waveform phasing, since the number of waveform cycles changes very little as we change the approximation used to compute them [8].

4.1. *Parameter estimation results.* Monte Carlo simulations of the inverse Fisher Matrix suggest that for a typical source, a 10 $M_\odot$ CO captured by a $10^6$ $M_\odot$ SMBH at SNR of 30, we expect to determine the CO and SMBH masses and the SMBH spin within fractional errors of $\sim 10^{-4}$, $10^{-3.5}$, and $10^{-4.5}$, respectively. We also expect to determine the location of the source on the sky and the SMBH spin orientation to within $10^{-3}$, $10^{-3.5}$ steradians, respectively (see [8] for results for white dwarfs and neutron star EMRIs). These estimates are broadly consistent with existing results in the literature [10]. However, our model is based on true geodesics of the Kerr space-time, and hence we have a more accurate frequency at plunge. Figure 1 illustrates the spread in errors that arise from randomisation of the extrinsic parameters for a typical source.

4.2. *Model-induced errors.* Our NK waveform includes conservative corrections to 2PN order that can be turned on and off to estimate their relevance for parameter estimation. This in turn allows us to compute the ratio $R$ of the model error that arises from omitting the conservative part of the SF to the Fisher Matrix error for each of the ten parameters in the model at SNR of 30. If $R \leq 1$ then the estimates obtained from a model that ignores the conservative piece should still be reliable, but if $R \gg 1$ then it is clear that the parameter estimates would not be trustworthy. We have found that for kludge model comparisons the ratio $R < 1$ for all parameters at most points in the parameter space for inspirals of white-dwarfs, neutron stars or black holes. In fact, less than 0.15% of points in the Monte Carlo runs satisfy $2 < R < 3$. Hence, including conservative corrections apparently might not be essential for accurate parameter determination, but we will certainly reduce the model errors if we include them up to 2PN order.

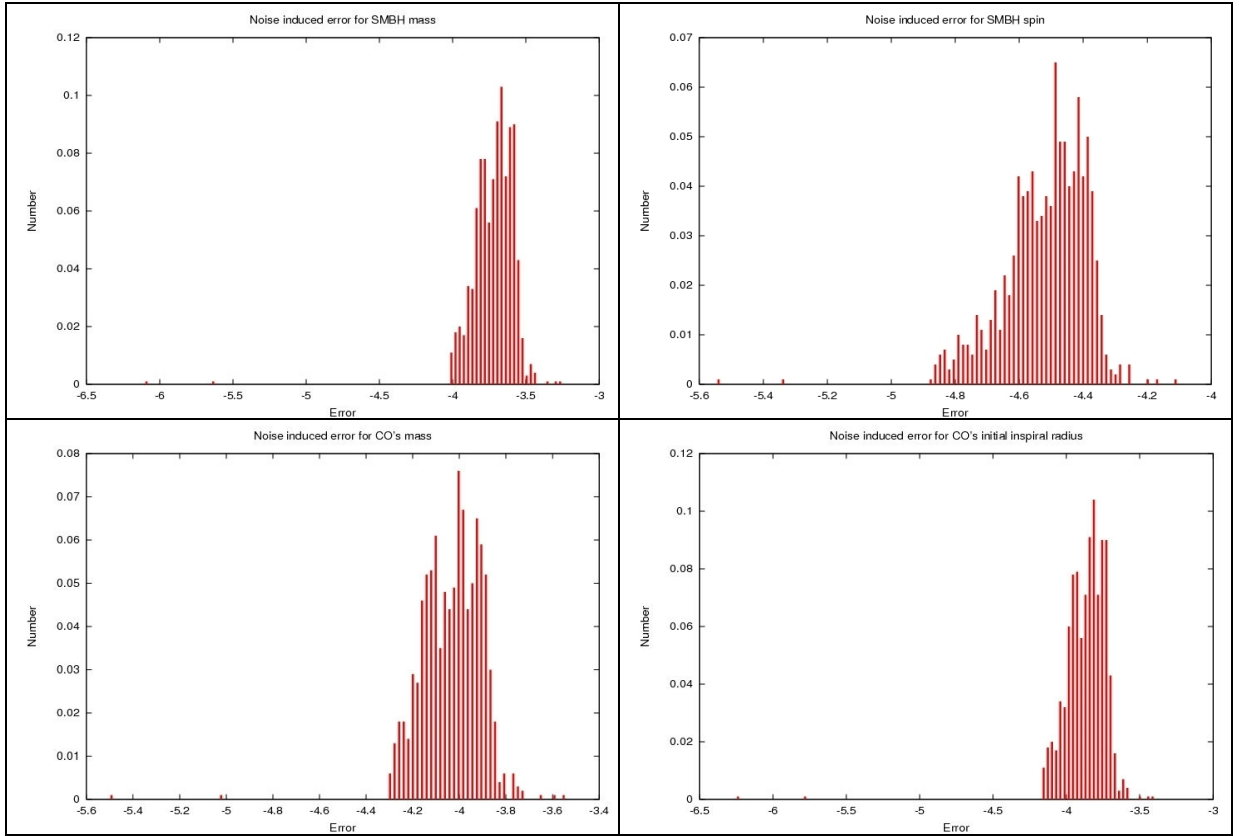

**Figure 1.** Distributions of errors in the intrinsic parameters as computed from the Monte Carlo simulations of the inverse Fisher Matrix. We show the system with $m=10\,M_\odot$, $M=10^6\,M_\odot$, SMBH's spin parameter q=0.9, and show the errors in log$(M)$, log$(q)$, log$(m)$ and log$(p_0)$, respectively.

*Model errors from first-order SF approximation.* We found that for white dwarfs and neutron stars inspirals, the ratio $R<1$ for practically every element of the parameter space. Indeed, less than 0.1% of points of the Monte Carlo runs satisfy $2<R<3$. On the other hand, we found that for black holes the ratios are typically larger than one. Part of the reason may be that the NK model was constructed by comparison to a weak-field PN expansion, but we are comparing it to fully accurate SF computations in the strong field. Therefore, it may be the case that the NK is itself not accurate enough in this particular regime. Nonetheless, these ratios are still manageably small.

## 5. Conclusions

We have developed an improved NK waveform model for circular-equatorial EMRIs by including conservative SF corrections up to 2PN order. We have found that the inclusion of conservative corrections has a relatively small impact on the waveform phasing. Hence, they are not essential for source detection, but it may be useful to include them for parameter estimation. At 2PN order, the NK model provides parameter estimation accuracy estimates that are broadly consistent with previous results [10]. For a typical source at SNR of 30, a LISA EMRI observation should be able to determine the CO and SMBH masses and the SMBH spin to within fractional errors of $\sim 10^{-4}$, $10^{-3.5}$, and $10^{-4.5}$, and determine the location of the source on the sky and the SMBH spin orientation to within $10^{-3}$, $10^{-3.5}$ steradians, respectively. Our model should be more reliable in the strong field regime as we have built it using a true Kerr geodesic, have included conservative corrections in a physically consistent way and have evolved the geodesic parameters using the best available radiative flux, $L_z$.

We have also assessed the importance of the first order conservative part and the second order radiative part of the SF for parameter estimation accuracy using the formalism introduced in [11].

We have found that the model errors that arise from omitting conservative SF terms are generally smaller than the parameter errors that arise from instrumental noise when the source has SNR=30. Furthermore, we have used SF calculations that include all first order terms of the SF, but nothing at higher order [5]. Comparing to these results using the NK allowed us to assess the relative importance of the first order conservative and second order radiative parts of the SF. We performed this exercise as these two contributions affect the orbital evolution in the same way. We found that the missing terms were not necessary for accurate parameter estimation for the inspirals of white dwarfs and neutron stars. However, the results for black holes are less conclusive as the model errors were typically a few times the expected parameter errors from instrumental noise. Part of the reason for this may have been that we derived the kludge corrections by comparison to PN expressions in the weak-field and then used them to test the SF results in the strong field.

These results, which are the first attempt to assess the necessity of including conservative corrections in templates for parameter estimation with LISA, suggest that search templates can ignore conservative corrections, but it may be necessary to follow up with more accurate templates, if available, in an area of the parameter space approximately ten times the error box predicted by the Fisher Matrix to get more precise parameter estimates. Additionally, the second order radiative piece of the SF may be as important as the first order conservative piece and it is not yet clear whether templates including both corrections will be available before LISA flies.

While we expect these results to be representative of the general case, we need to verify whether our conclusions hold true for EMRIs on orbits which are both eccentric and inclined to the equatorial plane.

**Acknowledgements**

EH acknowledges support from CONACyT, the Institute of Astronomy and Churchill College, Cambridge, and Stanford University. JG's work is supported by the Royal Society.